\documentclass[prb,twocolumn,showpacs]{revtex4}
\usepackage{amsmath}
\usepackage{dcolumn}
\usepackage{graphicx}
\usepackage{bm}

\begin{document}

\title{Insulator-to-metal phase transition in Yb-based Kondo insulators}
\author{Guang-Bin Li and Guang-Ming Zhang}
\affiliation{Department of Physics, Tsinghua University, Beijing 100084, China}
\author{Lu Yu}
\affiliation{Institute of Physics, Chinese Academy of Sciences, Beijing 100190, China;\\
Institute of Theoretical Physics, Chinese Academy of Sciences, Beijing
100190, China}
\date{\today }

\begin{abstract}
The periodic Anderson lattice model for the crystalline electric field (CEF)
split 4f quartet states is used to describe the Yb-based Kondo
insulators/semiconductors. In the slave-boson mean-field approximation, we
derive the hybridized quasiparticle bands, and find that decreasing the
hybridization difference of the two CEF quartets may induce an
insulator-to-metal phase transition. The resulting metallic phase has a hole
and an electron Fermi pockets. Such a phase transition may be realized
experimentally by applying pressure, reducing the difference in
hybridization of the two CEF quartets.
\end{abstract}

\maketitle

Kondo insulators or semiconductors, such as YbB$_{12}$, belong to strongly
correlated electron systems\cite{Aeppli1992,Riseborough2000}, in which the
conduction electrons hybridize with the localized 4f-electrons and the
strong Coulomb repulsion results in highly renormalized quasiparticle bands
with a small indirect energy gap\cite%
{Susaki1996,Okamura1998,Mignot2005,Nemkovski2007}. To study the
characteristic properties of these materials at low temperatures, some
experiments have been attempted to make the insulating gap vanish by
applying an external magnetic field\cite{Sugiyama1988,Jaime2000} or pressure%
\cite{Gabani2003}, leading to an insulator-metal phase transition. Such a
transition under the external magnetic field has been considered in the
previous studies\cite{Beach2003,Milat2004,Ohashi2004,Izumi2007}, however,
the microscopic mechanism for the pressure induced insulator-to-metal
transition remains far from being fully understood.

For the Yb-based Kondo insulators, an external pressure can affect the
hybridization between 5d band electrons and the more atomic-like 4f
electrons, giving rise to the intermediate valence behavior. The Yb valence
is directly related to the number of 4f-holes $n_{h}$ by $v=2+n_{h}$. At the
ambient pressure, $n_{h}$ spans a broad range between $0$ and $1$ in
Yb-based compounds, and the intermediate valence reflects the hybridization
of the energetically close Yb$^{2+}$ (4f$^{14}$) and Yb$^{3+}$ (4f$^{13}$)
configurations. The electronic configuration of Yb$^{3+}$ (4f$^{13}$) can be
regarded as a single hole in the 4f-shell, while the configuration of Yb$%
^{2+}$ (4f$^{14}$) corresponds to the closed 4f-shell. Taking into account
the much larger strength (1.3eV) of the spin-orbit coupling\cite%
{Martensson1982}, a $j=7/2$ f-hole state is split into a quartet and two
doublet states by the crystalline electric field (CEF) under the cubic
symmetry, which is the usual lattice structure of YbB$_{12}$. These two
doublets are almost degenerate and may be treated as a quasi-quartet. Thus,
a periodic Anderson lattice model with $U\rightarrow \infty $ for the CEF
split 4f states can be used to describe these Yb-based Kondo insulators or
semiconductors\cite{Akbari2009}.

It has been further pointed out that the anisotropic hybridizations of the
two CEF quartets play an important role in the formation of the two
dispersive spin resonances at the continuum threshold\cite{Akbari2009}, the
most salient features observed by inelastic neutron scattering experiments
in YbB$_{\text{12}}$ (Ref.\onlinecite{Mignot2005, Nemkovski2007}). Motivated
by this analysis, we further notice that, above a threshold of the CEF
splitting, decreasing the difference in the hybridization of the two CEF
quartets may cause an overlap between the middle lower and upper hybridized
quasiparticle bands, leading to an insulator-to-metal phase transition.
Experimentally, this phase transition can be realized by applying pressure,
reducing the difference in the hybridization of the two CEF quasi-quartets.
Such a pressure induced insulator-to-metal transition has been observed in
the Kondo insulator SmB$_{6}$, where the electrical resistivity has been
measured below $80$ K and under pressure between $1$ bar and $70$ kbar (Ref.%
\cite{Gabani2003}). Above the critical pressure $40$ kbar, a transition
occurs from a Kondo insulator to a metallic heavy fermion liquid and a
non-Fermi liquid behavior has been found\cite{Gabani2003}.

In this paper, we will carefully study the periodic Anderson model with $%
U\rightarrow \infty $ for the CEF split 4f quartet states. Using the
slave-boson mean-field approximation, we will derive four quasiparticle
bands resulting from the hybridization between the conduction electrons and
localized 4f-hole states, and find that decreasing the hybridization
difference of the two CEF quartets indeed induce an insulator-to-metal phase
transition. The resulting metallic phase has a hole and an electron Fermi
pockets. By including the Coulomb interaction between the localized and
conduction electrons, we discuss the possible instability of the resulting
metallic phase.

To describe the Yb-based Kondo insulators or semiconductors, the periodic
Anderson lattice model with $U\rightarrow \infty $ for the CEF split 4f
states has been introduced\cite{Akbari2009}%
\begin{align}
\mathcal{H}& =\sum_{\mathbf{k},\gamma }\epsilon _{\mathbf{k}}d_{\mathbf{k}%
\gamma }^{\dag }d_{\mathbf{k}\gamma }+\sum_{\mathbf{k},\gamma }(\varepsilon
_{f}+\Delta _{\gamma })f_{i\gamma }^{\dag }f_{i\gamma }  \notag \\
& +\frac{1}{\sqrt{\mathcal{N}}}\sum_{i,\mathbf{k},\gamma }(V_{\mathbf{k}%
\gamma }e^{i\mathbf{k\cdot R}_{i}}f_{i,\gamma }^{\dag }d_{\mathbf{k}\gamma
}b_{i}+h.c.),  \label{model}
\end{align}%
where the first term denotes the conduction electron band, the second term
stands for the binding energy of the 4f-hole, and $\Delta _{\gamma }$ ($%
\Delta _{1}=0,\Delta _{2}=\Delta $) is the CEF splitting energy for the two
quasi-quartets with $\gamma =(\Gamma ,m)$, where $\Gamma =1,2$ denotes the
quartets and $m=1-4$ represents the four-fold orbital degeneracy. Due to the
exclusion of the double occupancy, a projection has been implemented by
using the slave-boson representation\cite{Coleman1984}. Then the Yb$^{2+}$
(4f$^{14}$) configuration without a 4f-hole state can be accounted for by an
auxiliary boson state $b_{i}^{\dag }\left| 0\right\rangle $, while the Yb$%
^{3+}$ (4f$^{13}$) configuration with a 4f-hole state is represented by a
fermion state $f_{i\gamma }^{\dag }\left| 0\right\rangle $. The conduction
electrons hybridize with the $f$-hole at each lattice site in both quartets
with different strengths. At each lattice site the constraint $%
Q_{i}=b_{i}^{\dag }b_{i}+\sum_{\gamma }f_{i\gamma }^{\dag }f_{i\gamma }=1$
has to be enforced, and the total Hamiltonian is $\mathcal{H}%
+\sum_{i}\lambda _{i}(Q_{i}-1)$, where $\lambda _{i}$ is the Lagrange
multiplier.

Now the slave-boson mean-field approximation is performed by neglecting the
fluctuation of the Bose field $\langle b_{i}^{\dagger }\rangle =\left\langle
b_{i}\right\rangle =b$ and the site dependence of the local field $\lambda
_{i}=\lambda $. Within these approximations, the mean-field model
Hamiltonian can be written as
\begin{equation}
\mathcal{H}_{mf}=\sum_{\mathbf{k},\gamma }[\epsilon _{\mathbf{k}}d_{\mathbf{k%
}\gamma }^{\dag }d_{\mathbf{k}\gamma }+\tilde{\varepsilon}_{\gamma }f_{%
\mathbf{k}\gamma }^{\dag }f_{\mathbf{k}\gamma }+\tilde{V}_{\gamma }(d_{%
\mathbf{k}\gamma }^{\dag }f_{\mathbf{k}\gamma }+h.c.)]+\mathcal{N}\epsilon
_{0},
\end{equation}%
where $\tilde{\varepsilon}_{\gamma }=\varepsilon _{f}+\Delta _{\gamma
}+\lambda $ is the renormalized energy level of the localized states, $%
\tilde{V}_{\gamma }=bV_{\gamma }$, and $\epsilon _{0}=\lambda (b^{2}-1)$. It
should be noticed that the dependence of the hybridization strength on $%
\mathbf{k}$ has been neglected, \textit{i.e.}, $V_{\mathbf{k}\gamma
}=V_{\gamma }$. Furthermore, we will replace $V_{\gamma }$ by $V_{\Gamma }$
for simplicity. By performing the Bogoliubov transformation
\begin{equation}
\alpha _{\mathbf{k}\gamma }=\mu _{\mathbf{k\gamma }}d_{\mathbf{k}\gamma
}+\nu _{\mathbf{k}}f_{\mathbf{k}\gamma },\text{ }\beta _{\mathbf{k}\gamma
}=-\nu _{\mathbf{k\gamma }}d_{\mathbf{k}\gamma }+\mu _{\mathbf{k\gamma }}f_{%
\mathbf{k}\gamma },\text{\ }
\end{equation}%
we can diagonalize the quadratic Hamiltonian and obtain
\begin{equation}
\mathcal{H}_{MF}=\sum_{\mathbf{k},\gamma }\left( E_{\mathbf{k}\gamma
}^{+}\alpha _{\mathbf{k}\gamma }^{\dag }\alpha _{\mathbf{k}\gamma }+E_{%
\mathbf{k}\gamma }^{-}\beta _{\mathbf{k}\gamma }^{\dag }\beta _{\mathbf{k}%
\gamma }\right) ,
\end{equation}%
with four hybridized quasiparticle bands are

\begin{equation}
E_{\mathbf{k}\gamma }^{\pm }=\frac{1}{2}\left[ \epsilon _{\mathbf{k}}+\tilde{%
\varepsilon}_{\gamma }\pm \sqrt{(\epsilon _{\mathbf{k}}-\tilde{\varepsilon}%
_{\gamma })^{2}+4\tilde{V}_{\gamma }^{2}}\right] ,
\end{equation}%
while the Bogoliubov parameters $\mu _{\mathbf{k}\gamma }$ and $\nu _{%
\mathbf{k}\gamma }$ are given by%
\begin{equation}
\left(
\begin{array}{c}
\mu _{\mathbf{k}\gamma } \\
\nu _{\mathbf{k}\gamma }%
\end{array}%
\right) =\frac{1}{\sqrt{2}}\left[ 1\pm \frac{\epsilon _{\mathbf{k}}-\tilde{%
\varepsilon}_{\gamma }}{\sqrt{(\epsilon _{\mathbf{k}}-\tilde{\varepsilon}%
_{\gamma })^{2}+4\tilde{V}_{\gamma }^{2}}}\right] ^{1/2}.
\end{equation}%
These two parameters describe the contributions of the conduction electron
band and localized $f$-hole band to the hybridized quasiparticles,
respectively.

Moreover, the ground-state energy per site is given by
\begin{equation}
E_{g}=\frac{1}{\mathcal{N}}\sum_{\mathbf{k},\gamma }\left[ E_{\mathbf{k}%
\gamma }^{+}\theta (E_{\mathbf{k}\gamma }^{+})+E_{\mathbf{k}\gamma
}^{-}\theta (E_{\mathbf{k}\gamma }^{-})\right] +\epsilon _{0},
\end{equation}%
where $\theta (E_{\mathbf{k}\gamma }^{\pm })$ is the step function. The
chemical potential $\mu $ and the Lagrange multiplier $\lambda $ have to be
determined self-consistently according to the conservation of the total
number of particle per lattice site $n_{c}+n_{f}=2$. Depending on the
parameter values $\varepsilon _{f}$, $\Delta $, and$V_{\Gamma }$, the
variational parameters $b$ and $\lambda $ are also determined
self-consistently. From the hybridized quasiparticle band structure, the
ground state of the system can be an insulating state, where the two lower
bands are filled completely, leaving an indirect energy gap. As the $\mathbf{%
k}$ dependence in $E_{g}$ appears through the conduction electron energy $%
\epsilon _{\mathbf{k}}$, summations over $\mathbf{k}$ can be transformed
into an integral over energy $\epsilon $ in the interval $[-D,D]$. By
assuming a constant density of states, the ground-state energy is thus
evaluated as%
\begin{align}
E_{g}& =\frac{1}{8D}\sum_{\Gamma }\left\{ 4D\tilde{\varepsilon}_{\Gamma }-4%
\tilde{V}_{\Gamma }^{2}\ln \frac{\Lambda _{\Gamma }^{-}(D)+D-\tilde{%
\varepsilon}_{\Gamma }}{\Lambda _{\Gamma }^{+}(D)-D-\tilde{\varepsilon}%
_{\Gamma }}\right.  \notag \\
& -\left. [(D-\tilde{\varepsilon}_{\Gamma })\Lambda _{\Gamma }^{-}(D)+(D+%
\tilde{\varepsilon}_{\Gamma })\Lambda _{\Gamma }^{+}(D)]\right\} +\epsilon
_{0},
\end{align}%
where $\Lambda _{\Gamma }^{\pm }(x)=\sqrt{(x\pm \tilde{\varepsilon}_{\Gamma
})^{2}+4\tilde{V}_{\Gamma }^{2}}$, $\tilde{\varepsilon}_{2}=\tilde{%
\varepsilon}_{1}+\Delta ,$ and $\tilde{V}_{2}=bV_{2}=b(V_{1}+\delta V)$.
Minimizing the ground-state energy density with respect to $b$ and $\lambda $%
, respectively, we obtain the following self-consistent equations%
\begin{gather}
b^{2}=\frac{1}{4D}\sum_{\Gamma }\left[ \Lambda _{\Gamma }^{+}(D)-\Lambda
_{\Gamma }^{-}(D)\right] ,  \notag \\
\lambda =\frac{1}{2D}\sum_{\Gamma }V_{\Gamma }^{2}\ln \frac{\Lambda _{\Gamma
}^{-}(D)+D-\tilde{\varepsilon}_{\Gamma }}{\Lambda _{\Gamma }^{+}(D)-D-\tilde{%
\varepsilon}_{\Gamma }}.  \label{insulator}
\end{gather}

However, we notice that there exists another possible structure of the
quasiparticle bands, where the chemical potential $\mu $ cuts through the
two middle hybridized quasiparticle bands $E_{\mathbf{k}1}^{+}$ and $E_{%
\mathbf{k}2}^{-}$ at $\xi _{1}$ and $\xi _{2}$, respectively. Both these
energy parameters are determined by the equation $E_{\mathbf{k}1}^{+}=E_{%
\mathbf{k}2}^{-}=\mu $. From the condition of the total number of particles
per lattice site $n_{c}+n_{f}=2$, we can derive the result $\xi _{1}=-\xi
_{2}\equiv -\xi $ and
\begin{equation}
2\xi +\Delta =\Lambda _{1}^{+}(\xi )+\Lambda _{2}^{-}(\xi ).
\label{order-para}
\end{equation}%
Here $\xi $ can be used to characterize the insulator-to-metal transition.
When $0<\xi <D$, the ground state should be metallic, while for $\xi =D$ the
ground state corresponds to a critical point. The corresponding ground-state
energy density in the metallic phase is thus expressed as
\begin{align}
E_{g}& =\frac{1}{4D}[(3D-\xi )\tilde{\varepsilon}_{1}+(D+\xi )\tilde{%
\varepsilon}_{2}+\xi ^{2}-D^{2}]  \notag \\
& +\frac{\tilde{V}_{1}^{2}}{2D}\ln \frac{\Lambda _{1}^{+}(\xi )-\xi -\tilde{%
\varepsilon}_{1}}{\Lambda _{1}^{-}(D)+D-\tilde{\varepsilon}_{1}}-\frac{%
\tilde{V}_{2}^{2}}{2D}\ln \frac{\Lambda _{2}^{-}(\xi )+\xi -\tilde{%
\varepsilon}_{2}}{\Lambda _{2}^{+}(D)-D-\tilde{\varepsilon}_{2}}  \notag \\
& -\frac{1}{8D}[(\xi +\tilde{\varepsilon}_{1})\Lambda _{1}^{+}(\xi )+(D-%
\tilde{\varepsilon}_{1})\Lambda _{1}^{-}(D)]  \notag \\
& -\frac{1}{8D}[(\xi -\tilde{\varepsilon}_{2})\Lambda _{2}^{-}(\xi )+(D+%
\tilde{\varepsilon}_{2})\Lambda _{2}^{+}(D)]+\epsilon _{0}.
\end{align}%
By minimizing $E_{g}$ with respect to $b$ and $\lambda $, the corresponding
self-consistent equations can be deduced to
\begin{gather}
b^{2}=\frac{1}{4D}[\Lambda _{1}^{+}(\xi )-\Lambda _{1}^{-}(D)-\Lambda
_{2}^{-}(\xi )+\Lambda _{2}^{+}(D)],  \notag \\
\lambda =\frac{V_{1}^{2}}{2D}\ln \frac{\Lambda _{1}^{-}(D)+D-\tilde{%
\varepsilon}_{1}}{\Lambda _{1}^{+}(\xi )-\xi -\tilde{\varepsilon}_{1}}+\frac{%
V_{2}^{2}}{2D}\ln \frac{\Lambda _{2}^{-}(\xi )+\xi -\tilde{\varepsilon}_{2}}{%
\Lambda _{2}^{+}(D)-D-\tilde{\varepsilon}_{2}}.  \label{metal}
\end{gather}

\begin{figure}[tbp]
\begin{center}
\includegraphics[width=6.5cm,angle=-90]{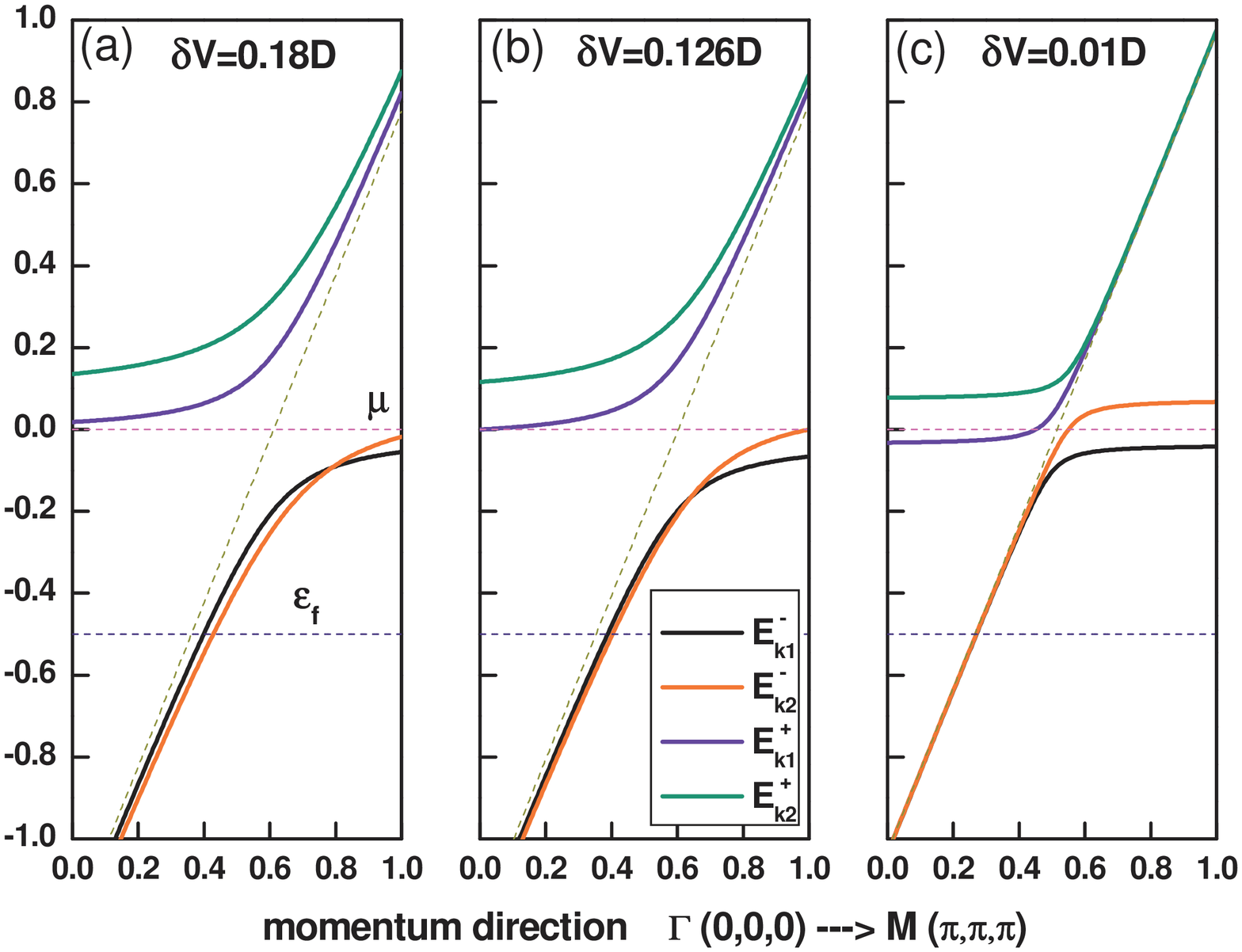}
\end{center}
\caption{(color online). The quasiparticle band structure for $V_{1}=0.4D$, $%
\protect\epsilon _{f}=-0.5D$, and $\Delta =0.1D$. (a) Insulator phase, (b)
Critical point, (c) Metallic phase.}
\label{spectra}
\end{figure}

In order to deduce the ground state phase diagram, we should first
numerically solve Eq.(\ref{insulator}) for the insulating phase and Eqs.(\ref%
{order-para}) and (\ref{metal}) for the metallic phase, respectively. The
hybridized quasiparticle band energy versus the momentum along the diagonal
direction $\Gamma $ $(0,0,0)->M$ $(\pi ,\pi ,\pi )$ are plotted in Fig.\ref%
{spectra} with $V_{1}=0.4D$, $\epsilon _{f}=-0.5D$, and $\Delta =0.1D$ for
three different values of $\delta V$. As shown in Fig.\ref{spectra}(a) for $%
\delta V=0.18D$, there opens an indirect gap between the middle upper and
lower bands, corresponding to an insulating phase. In Fig.\ref{spectra}(b)
for $\delta V=0.126D$, the middle upper and lower bands just meet at the
chemical potential, corresponding the critical point of the transition.
Since we have $\xi =D$ at the critical point, the ground-state energies of
the metallic and insulating phases are equal. So the insulator-metal
transition is a continuous second-order phase transition. Finally, in Fig.%
\ref{spectra}(c) for $\delta V=0.01D$, the middle lower and upper bands
overlap, and the chemical potential cuts through these two bands, which
corresponds to the metallic phase.

\begin{figure}[tbp]
\begin{center}
\includegraphics[width=3cm,angle=-90]{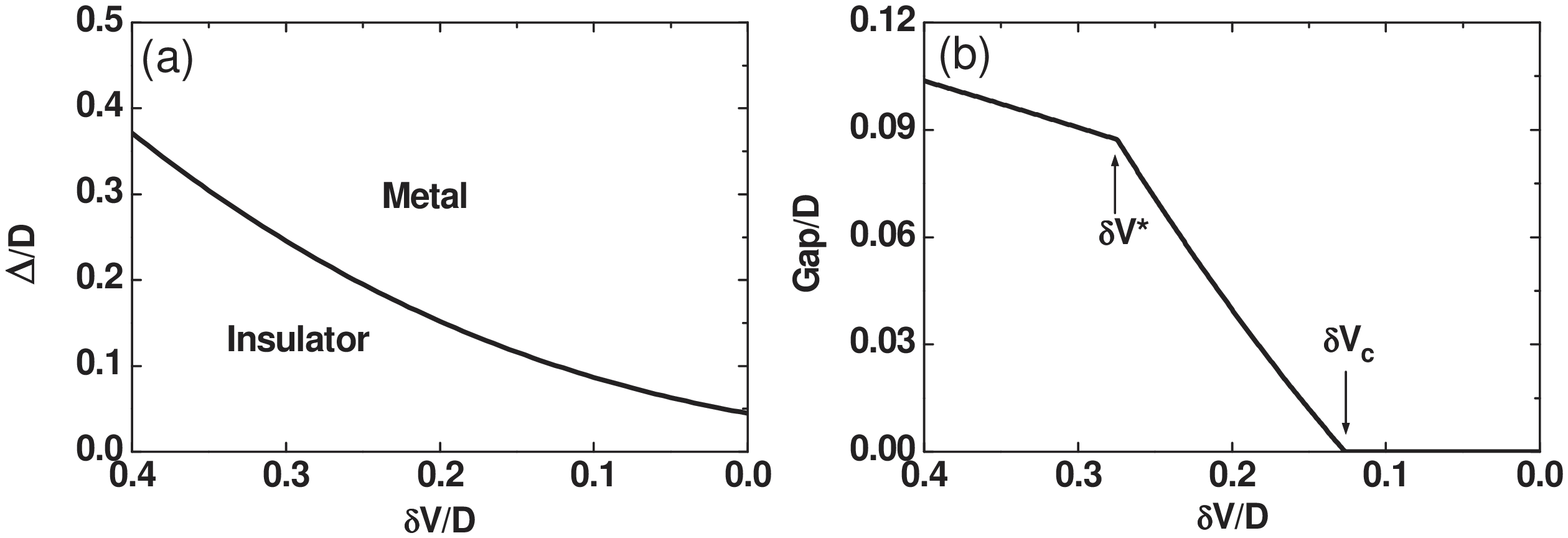}
\end{center}
\caption{(a) Ground state phase diagram for $V_{1}=0.4D$ and $\protect%
\epsilon _{f}=-0.5D$; (b) Indirect energy gap as a function of the
hybridization difference of the two CEF quartets $\protect\delta V$ for $%
\Delta =0.1D$.}
\label{phasediagram}
\end{figure}

The critical condition under which the insulator-metal transition occurs can
be determined from Eq.(\ref{order-para}) and Eq.(\ref{metal}) by setting $%
\xi =D$. Then the ground-state phase diagram can constructed for $V_{1}=0.4D$
and $\epsilon _{f}=-0.5D$ and is shown in Fig.\ref{phasediagram}(a). Clearly
there exists a threshold of the CEF splitting energy $\Delta _{c}$, and only
when $\Delta >\Delta _{c}$ the insulator-to-metal phase transition occurs by
turning the difference in hybridization of the two CEF quasi-quartets. The
change of the indirect gap is another evidence to characterize the
insulator-to-metal phase transition, and can be also calculated and
displayed in Fig.\ref{phasediagram}(b) for $\Delta =0.1D$. It shows that the
indirect quasiparticle gap decreases almost linearly with decreasing the
hybridization difference of the two CEF quartets, and this energy gap
finally vanishes at $\delta V_{c}$. There is another critical value $\delta
V^{\ast }$, where the top energy levels of the two lower quasiparticle bands
interchange with each other around the Brillouin zone boundary. Then the
indirect energy gap has a cusp.

Actually, such an insulator-to-metal phase transition can be realized
experimentally. There exists a strong CEF splitting estimated in YbB$_{12}$,
and we believe that increasing pressure can continuously reduce the
difference in hybridization of the two CEF quasi-quartets. So below the
critical value $\delta V_{c}$, YbB$_{12}$ is an insulator with an indirect
gap as observed in experiments \cite%
{Susaki1996,Okamura1998,Mignot2005,Nemkovski2007}, while above this critical
value $\delta V_{c}$ this material can transform into a heavy electron metal
with an enhanced effective mass due to the presence of heavy charge
carriers. Thus, our theory may provide a general microscopic mechanism of
the pressure induced insulator-to-metal transition in Yb-based Kondo
insulators/semiconductors.

\begin{figure}[tbp]
\begin{center}
\includegraphics[angle=-90,width=8.5cm]{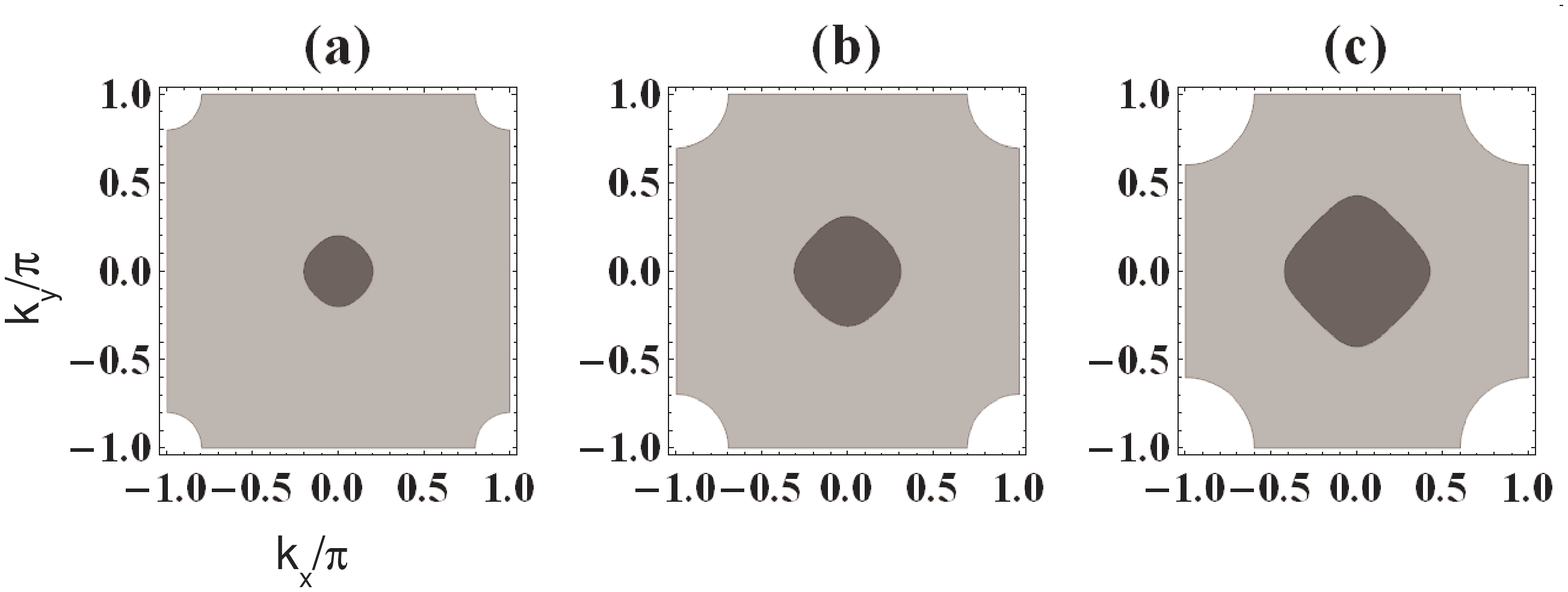}
\end{center}
\caption{(color online). Fermi surface structure of the two-dimensional
model Hamiltonian on a square lattice with $V_{1}=0.4D$, $\Delta =0.1D$, and
$\protect\epsilon _{f}=-0.5D$. (a) $\protect\delta V=0.08D$, (b) $\protect%
\delta V=0.05D$ and (c) $\protect\delta V=0.01D$.}
\label{fermisurface}
\end{figure}

Since a constant density of states for the conduction electron band was
assumed in the above slave-boson mean-field calculation, the obtained
results are independent of the dimensionality of the model. In order to see
the special Fermi surface structure of the metallic phase, the model
Hamiltonian Eq.(\ref{model}) is redefined on a two-dimensional square
lattice system with the conduction electron band%
\begin{equation}
\epsilon _{\mathbf{k}}=-2t(\cos k_{x}+\cos k_{y})+4t^{\prime }\cos k_{x}\cos
k_{y},
\end{equation}%
where $t$ denotes the nearest neighbor hopping and $t^{\prime }$ denotes the
next-nearest neighbor hopping. Then the same slave-boson mean field
calculation can be performed, and the insulator-to-metal phase transition
also takes place for a set of parameters $V_{1}=0.4D$, $\epsilon _{f}=-0.5D$%
, $t=0.25D$, and $t^{\prime }=0.3t$ when decreasing the parameter $\delta V$%
. \ In the metallic phase, we have calculated the corresponding Fermi
surface structure shown in Fig.\ref{fermisurface}. There exist two Fermi
pockets: one electron-like in the center of the Brillouin zone and one
hole-like in the corners of the Brillouin zone. These two Fermi pockets have
exactly the same area in the Brillouin zone. Such a heavy electron metal
corresponds to a semi-metal. Decreasing the hybridization difference $\delta
V$ below the critical value, the sizes of the electron and hole Fermi
pockets become larger and larger, as displayed in Fig.\ref{fermisurface}.

For the three dimensional Yb-based Kondo insulators/semiconductors, the
Fermi surface of the resulting metallic phase should still be given by a
hole and an electron pockets. As the temperature is lowered enough, some
instabilities may further appear. Due to the presence of the strong mixed
valence effect in such systems, the additional on-site Coulomb interaction
between the conduction electrons and localized $f$-hole should be taken into
account. In the slave-boson representation, it is given by
\begin{equation}
\mathcal{H}_{I}=U_{fc}\sum_{i}\sum_{\gamma \gamma ^{\prime }}f_{i\gamma
}^{\dag }f_{i\gamma }d_{i\gamma ^{\prime }}^{\dag }d_{i\gamma ^{\prime }}.
\end{equation}%
When the coupling strength $U_{fc}$ is assumed to be small, we can rewrite
this additional interaction in terms of the hybridized quasiparticles as%
\begin{gather}
\mathcal{H}_{I}=\frac{U_{fc}}{\mathcal{N}}\sum_{\mathbf{\mathbf{k}}_{1}%
\mathbf{k}_{2}\mathbf{\mathbf{k}}_{3}\mathbf{\mathbf{k}}_{4}}\left( \nu _{%
\mathbf{k}_{1}1}\mu _{\mathbf{k}_{2}1}\nu _{\mathbf{\mathbf{k}}_{3}1}\mu _{%
\mathbf{\mathbf{k}}_{4}1}\alpha _{\mathbf{\mathbf{k}}_{3}1}^{\dag }\alpha _{%
\mathbf{k}_{1}1}\alpha _{\mathbf{\mathbf{k}}_{4}1}^{\dag }\alpha _{\mathbf{k}%
_{2}1}\right. \text{ }  \notag \\
+\mu _{\mathbf{k}_{1}2}\nu _{\mathbf{k}_{2}2}\mu _{\mathbf{\mathbf{k}}%
_{3}2}\nu _{\mathbf{\mathbf{k}}_{4}2}\beta _{\mathbf{\mathbf{k}}_{3}2}^{\dag
}\beta _{\mathbf{k}_{1}2}\beta _{\mathbf{\mathbf{k}}_{4}2}^{\dag }\beta _{%
\mathbf{k},2}  \notag \\
+\nu _{\mathbf{k}_{1}1}\nu _{\mathbf{k}_{2}2}\nu _{\mathbf{\mathbf{k}}%
_{3}1}\nu _{\mathbf{\mathbf{k}}_{4}2}\alpha _{\mathbf{\mathbf{k}}%
_{3}1}^{\dag }\alpha _{\mathbf{k}_{1}1}\beta _{\mathbf{\mathbf{k}}%
_{4}2}^{\dag }\beta _{\mathbf{k}_{2}2}  \notag \\
\text{ \ \ \ \ \ }+\left. \mu _{\mathbf{k}_{1}2}\mu _{\mathbf{k}_{2}1}\mu _{%
\mathbf{\mathbf{k}}_{3}2}\mu _{\mathbf{\mathbf{k}}_{4}1}\beta _{\mathbf{%
\mathbf{k}}_{3}2}^{\dag }\beta _{\mathbf{k}_{1}2}\alpha _{\mathbf{\mathbf{k}}%
_{4}1}^{\dag }\alpha _{\mathbf{k}_{2}1}\right) ,
\end{gather}%
where $\mathbf{\mathbf{k}}_{1}+\mathbf{k}_{2}\mathbf{=\mathbf{k}}_{3}+%
\mathbf{\mathbf{k}}_{4}$ should be satisfied and \textit{only} the two
quasiparticle bands, crossing the Fermi energy, have been taken into
account. $\alpha _{\mathbf{k},1}$ and $\alpha _{\mathbf{k},1}^{\dagger }$
are defined on the electron Fermi pocket, while $\beta _{\mathbf{k},2}$ and $%
\alpha _{\mathbf{k},2}^{\dagger }$ are defined on the hole Fermi pocket.
Among these residual quasiparticle interactions, the first two terms
represent the intra-pocket scatterings with a small momentum transfer, while
the last two terms correspond to the inter-pocket scatterings with a large
momentum transfer.

According to the recent renormalization group analysis for a two-band
interacting model with electron and hole Fermi pockets\cite{Chubukov2008},
the inter-pocket quasiparticle interactions will determine the possible
instabilities at low temperatures. When we set $\mathbf{q}$ as a small
momentum and $\mathbf{Q}$ as a large momentum which is the distance between
the centers of two Fermi pockets, then inter-pocket quasiparticle
interactions can be approximated as%
\begin{eqnarray}
&&-\frac{U_{fc}}{\mathcal{N}}\sum_{\mathbf{qq}^{\prime }}\text{ }(\mu _{%
\mathbf{q}1}\mu _{\mathbf{q}^{\prime }1}\mu _{\mathbf{q}2}\mu _{\mathbf{q}%
^{\prime }2}+\nu _{\mathbf{q}1}\nu _{\mathbf{q}^{\prime }1}\nu _{\mathbf{q}%
2}\nu _{\mathbf{q}^{\prime }2})  \notag \\
&&\text{ \ \ \ \ \ \ \ \ }\times \alpha _{\mathbf{q\mathbf{,}}1}^{\dag
}\beta _{\mathbf{Q+q},2}\beta _{\mathbf{Q+q}^{\prime }2}^{\dag }\alpha _{%
\mathbf{q}^{\prime },1}.
\end{eqnarray}%
If there is a strong nesting between the hole and electron Fermi pockets,
this inter-pocket repulsive interaction will further induce a particle-hole
pairing instability, corresponding to an orbital-density wave ordering. The
corresponding order parameter is given by $\langle \alpha _{\mathbf{q}%
1}^{\dag }\beta _{\mathbf{Q+q}2}\rangle $ or $\langle \beta _{\mathbf{Q+q}%
^{\prime }2}^{\dag }\alpha _{\mathbf{q}^{\prime }1}\rangle $. Such a new
type of ordering in heavy fermion materials will be discussed in our further
investigations.

In conclusion, we have studied the Yb-based Kondo insulators with a strong
CEF splitting in the framework of the periodic Anderson lattice model by
using the slave-boson mean-field approximation. The obtained ground-state
phase diagram and the indirect gap have demonstrated that a second-order
insulator-to-metal transition occurs via reducing the hybridization
difference of the two CEF quasi-quartets. Our theory provides a general
microscopic mechanism of the pressure induced insulator-to-metal transition,
because increasing the external pressure can effectively reduce the
anisotropy of the hybridization strengths of the two CEF quartets
experimentally. The resulting metallic phase has a hole and an electron
Fermi pockets, which may exhibit an instability of an orbital-density wave
ordering at low temperatures when the inter-pocket quasiparticle residual
interactions are taken into account. These theoretical results are certainly
needed to be confirmed experimentally in the future.

The authors would like to thank Dung-Hai Lee for his stimulating discussions
and Yu Liu for his helps in the numerical calculations. This work is
partially supported by NSF-China and the National Program for Basic Research
of MOST, China.


\begin{thebibliography}{99}
\bibitem{Aeppli1992} G. Aeppli and Z. Fisk, Comments Cond. Mat. Phys.
\textbf{16}, 155-165 (1992).

\bibitem{Riseborough2000} P. S. Riseborough, Adv. Phys. \textbf{49}, 257
(2000).

\bibitem{Susaki1996} T. Susaki, \textit{et. al.}, Phys. Rev. Lett. \textbf{77%
}, 4269 (1996).

\bibitem{Okamura1998} H. Okamura, S. Kimura, H. Shinozaki, T. Nanba, F. Iga,
N. Shimizu, and T. Takabatake, Phys. Rev. B \textbf{58}, 7496 (1998).

\bibitem{Mignot2005} J. M. Mignot, P. A. Alekseev, K. S. Nemkovski, L. P.
Regnault, F. Iga, and T. Takabatake, Phys. Rev. Lett. \textbf{94}, 247204
(2005).

\bibitem{Nemkovski2007} K. S. Nemkovski,\textit{et al.}, Phys. Rev. Lett.
\textbf{99}, 137204 (2007).

\bibitem{Sugiyama1988} K. Sugiyama, F. Iga, M. Kasaya, T. Kasuya, and M.
Date, J. Phys. Soc. Jpn. \textbf{57}, 3946 (1988).

\bibitem{Jaime2000} M. Jaime, R. Movshovich, G. R. Stewart, W. P. Beyermann,
M. G. Berisso, M. F. Hundley, P. C. Canfield, and J. L. Sarrao, Nature
\textbf{405}, 160 (2000).

\bibitem{Gabani2003} S. Gab\'{a}ni, E. Bauer, S. Berger, K. Flachbart, Y.
Paderno, C. Paul, V. Pavl\'{\i}k, and N. Shitsevalova, Phys. Rev. B \textbf{%
67}, 172406 (2003).

\bibitem{Beach2003} K. S. D. Beach, P. A. Lee, and P. Monthoux, Phys. Rev.
Lett. \textbf{92}, 026401 (2004).

\bibitem{Milat2004} I. Milat, F. Assaad, and M. Sigrist, Eur. Phys. J. B
\textbf{38}, 571 (2004).

\bibitem{Ohashi2004} T. Ohashi, A. Koga, S. Suga, and N. Kawakami, Phys.
Rev. B \textbf{70}, 245104 (2004).

\bibitem{Izumi2007} T. Izumi, Y. Imai, and T. Saso, J. Phys. Soc. Jpn.
\textbf{76}, 4715 (2007).

\bibitem{Martensson1982} N. Martensson, B. Reihl, R. A. Pollak, F.
Holtzheng, G. Kaindle, Phys. Rev. B \textbf{25}, 6522 (1982).

\bibitem{Akbari2009} A. Akbari, P. Thalmeier and P. Fulde, Phys. Rev. Lett.
\textbf{102}, 106402 (2009).

\bibitem{Coleman1984} P. Coleman, Phys. Rev. B \textbf{29}, 3035 (1984).

\bibitem{Chubukov2008} A. V. Chubukov, D. Efremov, and I. Eremin, Phys. Rev.
B \textbf{78}, 134512 (2008).
\end{thebibliography}
\end{document}